\documentclass[reprint,showpacs,preprintnumbers,nofootinbib,amsmath,amssymb,aps,prd,floatfix]{revtex4-1}
\pdfoutput=1
\usepackage{graphicx}
\usepackage{dcolumn}
\usepackage{bm}
\usepackage{url}
\usepackage[ddmmyy,24hr]{datetime}
\usepackage{varwidth}
\usepackage{multirow}
\begin{document}

\preprint{\begin{tabular}{l}
JHEP 1311 (2013) 211
\\
\texttt{arXiv:1309.3192 [hep-ph]}
\end{tabular}}

\title{Light Sterile Neutrinos in Cosmology and Short-Baseline Oscillation Experiments}

\author{S. Gariazzo}
\affiliation{Department of Physics, University of Torino,
and
INFN, Sezione di Torino,
Via P. Giuria 1, I--10125 Torino, Italy}

\author{C. Giunti}
\affiliation{INFN, Sezione di Torino,
and
Department of Physics, University of Torino,
Via P. Giuria 1, I--10125 Torino, Italy}

\author{M. Laveder}
\affiliation{Dipartimento di Fisica e Astronomia ``G. Galilei'', Universit\`a di Padova,
and
INFN, Sezione di Padova,
Via F. Marzolo 8, I--35131 Padova, Italy}


\begin{abstract}
We analyze the most recent cosmological data,
including Planck,
taking into account the possible existence of a sterile neutrino with a mass at the eV scale
indicated by short-baseline neutrino oscillations data in the 3+1 framework.
We show that the contribution of local measurements of the Hubble constant
induces an increase of the value of the effective number of relativistic degrees of freedom
above the Standard Model value,
giving an indication in favor of the existence of sterile neutrinos
and their contribution to dark radiation.
Furthermore,
the measurements of the local galaxy cluster mass distribution
favor the existence of sterile neutrinos with eV-scale masses,
in agreement with short-baseline neutrino oscillations data.
In this case there is no tension between
cosmological and short-baseline neutrino oscillations data,
but the contribution of the sterile neutrino
to the effective number of relativistic degrees of freedom
is likely to be smaller than one.
Considering the
Dodelson-Widrow and thermal models
for the statistical cosmological distribution of sterile neutrinos,
we found that in the
Dodelson-Widrow model
there is a slightly better compatibility between
cosmological and short-baseline neutrino oscillations data
and the required suppression
of the production of sterile neutrinos in the early Universe
is slightly smaller.
\end{abstract}


\maketitle

\section{Introduction}
\label{Introduction}

The recent results of the Planck experiment
\cite{1303.5062,1303.5076}
generated lively discussions
\cite{1301.3119,1303.6267,1303.6270,1304.5243,1304.5981,1304.6217,1305.1971,1306.6766,1307.2904,1307.7715,1308.3255,1308.5870}
on the value of the effective number of relativistic degrees of freedom
$N_{\text{eff}}$
before photon decoupling
(see \cite{Giunti:2007ry,1212.6154,Lesgourgues-Mangano-Miele-Pastor-2013}),
which gives the energy density of radiation $\rho_R$ through the relation
\begin{equation}\label{eq:raddensitygen}
\rho_R=\left[1+\frac{7}{8}\left(\frac{4}{11}\right)^{4/3} N_{\text{eff}}\right]\rho_\gamma ,
\end{equation}
where
$\rho_\gamma$
is the photon energy density.
Since the value of $N_{\text{eff}}$ in the Standard Model (SM) is $N_{\text{eff}}^{\text{SM}}=3.046$
\cite{astro-ph/0111408,hep-ph/0506164},
a positive measurement of
$\Delta N_{\text{eff}} = N_{\text{eff}} - N_{\text{eff}}^{\text{SM}}$
may be a signal that the radiation content of the universe was due not only to photons and SM neutrinos,
but also to some additional light particle called generically ``dark radiation''.

In this paper we consider the possiblity that the dark radiation is made of the light sterile neutrinos
(see \cite{0704.1800,1204.5379,1302.1102,1303.6912})
whose existence is indicated by recent results of short-baseline (SBL) neutrino oscillation experiments
\cite{1006.3244,1101.2755,1207.4765,1210.5715,1212.3805,1303.3011,1308.5288}.
In particular, we consider the simplest possibility of a 3+1 scheme,
in which the three active flavor neutrinos
$\nu_{e}$,
$\nu_{\mu}$,
$\nu_{\tau}$,
are mainly composed of
three very light neutrinos
$\nu_{1}$,
$\nu_{2}$,
$\nu_{3}$,
with masses much smaller than 1 eV
and there is a sterile neutrino
$\nu_{s}$
which is mainly composed of a new massive neutrino
$\nu_{4}$
with mass
$m_{4} \sim 1 \, \text{eV}$.

The problem of the determination of $N_{\text{eff}}$
from cosmological data is related to that of the Hubble constant $H_0$,
because these two quantities are positively correlated in the analysis of the data
(see Refs.~\cite{1104.2333,1307.0637}).
Since dedicated local astrophysical experiments
obtained values of $H_0$ which are larger than that obtained
by the Planck collaboration
from the analysis of
cosmological data alone
\cite{1303.5076},
there is an indication that $N_{\text{eff}}$ may be larger than
the SM value.
We discuss this problem in Section~\ref{Cosmological},
where we present the results of a fit of
cosmological data with a prior on $H_0$ determined by the weighted average
of the local astrophysical measurements.

Since the neutrino oscillation explanation of SBL data requires the existence of a
massive neutrino at the eV scale,
we discuss also the bounds on the effective sterile neutrino mass
$m_{s}^{\text{eff}}$
defined by the Planck collaboration as \cite{1303.5076}
\begin{equation}\label{eq:meffsteriledef}
m_{s}^{\text{eff}} = \left( 94.1 \, \text{eV} \right) \Omega_{s} h^2
\,,
\end{equation}
where
$\Omega_{s} = \rho_{s} / \rho_{c}$
and
$h$ is the reduced Hubble constant, such that
$H_{0} = 100 \, h \, \text{km} \, \text{s}^{-1} \, \text{Mpc}^{-1}$.
Here
$\rho_{s}$ is the current energy density of $\nu_{s} \simeq \nu_{4}$
with mass
$m_{s} \simeq m_{4} \sim 1 \, \text{eV}$
and
$\rho_{c}$ is the current critical density.
The constant in Eq.~(\ref{eq:meffsteriledef}) refers to a Fermi-Dirac distribution
with the standard neutrino temperature
$T_\nu = (4/11)^{1/3} T_\gamma$.
We consider the two cases discussed
by the Planck collaboration \cite{1303.5076}
(see also \cite{0812.2249}):

\begin{description}

\item[Thermal (TH) model]
The sterile neutrino has a Fermi-Dirac distribution
$f_{s}(E) = (e^{E/T_{s}}+1)^{-1}$
with a temperature $T_{s}$ which is
different from the temperature $T_{\nu}$ of the active neutrinos,
leading to
\begin{equation}
m_{s}^{\text{eff}}
=
(T_{s}/T_{\nu})^3
m_{s}
=
(\Delta N_{\text{eff}})^{3/4}
m_{s}
\,.
\label{THmeff}
\end{equation}

\item[Dodelson-Widrow (DW) model]
The sterile neutrino has a Fermi-Dirac distribution
$f_{s}(E) = \chi_{s}/(e^{E/T_{\nu}}+1)$,
with the same temperature $T_{\nu}$ as the active neutrinos
but multiplied by a constant scale factor
$\chi_{s}$
\cite{hep-ph/9303287}.
In this case
\begin{equation}
m_{s}^{\text{eff}}
=
\chi_{s}
\,
m_{s}
=
\Delta N_{\text{eff}}
\,
m_{s}
\,.
\label{DWmeff}
\end{equation}

\end{description}

A further important problem is the
compatibility of the cosmological bounds on
$N_{\text{eff}}$
and
$m_{s}^{\text{eff}}$
with the active-sterile neutrino mixing required to
fit SBL oscillation data.
The stringent bounds on
$N_{\text{eff}}$
and
$m_{s}^{\text{eff}}$
presented in Ref.~\cite{1303.5076}
by the Planck collaboration imply
\cite{1303.5368}
that
the production of sterile neutrinos in the early Universe
is suppressed by some non-standard mechanism,
as, for example,
a large lepton asymmetry
\cite{1204.5861,1206.1046,1302.1200,1302.7279}.
In this paper we adopt a phenomenological approach similar to that in
Refs.~\cite{1207.6515,1302.6720,1303.4654}:
we use the results of the fit of SBL neutrino oscillation data
as a prior for the analysis of cosmological data.
In this way,
in Section~\ref{SBL}
we derive the combined constraints on
$N_{\text{eff}}$
and
$m_{s}^{\text{eff}}$
and the related constraints on $H_0$ and $m_{s}$.

\begin{table}[b]
\caption{\label{tab:H0}
Global best-fit value $H_0^{\text{gbf}}$,
marginal best-fit $H_0^{\text{mbf}}\pm1\sigma$ (68.27\%) and
$2\sigma$ (95.45\%) limits for $H_0$
obtained from the analysis of the indicated data sets.}
\begin{center}
\begin{tabular}{clccc}
\multicolumn{2}{c}{data} & $H_0^{\text{gbf}}$ & $H_0^{\text{mbf}}\pm1\sigma$ & $2\sigma$ \\
\hline
\multirow{3}{*}{
\begin{varwidth}{4em}
\center no\\SBL\\prior
\end{varwidth}}
	&CMB+$H_0$		&$73.6$	&$72.7^{+1.9}_{-1.7}$	&$69.0 \div 76.3$ \\
	&CMB+$H_0$+BAO		&$71.1$	&$71.5^{+1.4}_{-1.4}$	&$68.7 \div 74.4$ \\
	&CMB+$H_0$+BAO+LGC	&$71.1$	&$70.4^{+1.5}_{-1.3}$	&$68.1 \div 73.5$ \\
\hline
\multirow{4}{*}{
\begin{varwidth}{4em}
\center TH\\SBL\\prior
\end{varwidth}}
	&CMB			&$66.8$	&$66.6^{+1.1}_{-1.2}$	&$64.3 \div 68.9$ \\
	&CMB+$H_0$		&$68.7$	&$68.7^{+1.0}_{-1.1}$	&$66.5 \div 70.7$ \\
	&CMB+$H_0$+BAO		&$68.7$	&$68.8^{+0.8}_{-0.7}$	&$67.3 \div 70.4$ \\
	&CMB+$H_0$+BAO+LGC	&$69.1$	&$69.3^{+0.6}_{-0.6}$	&$68.1 \div 70.6$ \\
\hline
\multirow{4}{*}{
\begin{varwidth}{4em}
\center DW\\SBL\\prior
\end{varwidth}}
	&CMB			&$66.5$	&$66.9^{+1.2}_{-1.3}$	&$64.6 \div 69.4$ \\
	&CMB+$H_0$		&$68.1$	&$68.9^{+1.1}_{-1.0}$	&$66.9 \div 71.0$ \\
	&CMB+$H_0$+BAO		&$69.3$	&$69.1^{+0.8}_{-0.8}$	&$67.6 \div 70.6$ \\
	&CMB+$H_0$+BAO+LGC	&$69.5$	&$69.7^{+0.7}_{-0.5}$	&$68.6 \div 71.0$ \\
\hline
\end{tabular}
\end{center}
\end{table}

\begin{table}[b]
\caption{\label{tab:neff}
Global best-fit value $N_{\text{eff}}^{\text{gbf}}$,
marginal best-fit $N_{\text{eff}}^{\text{mbf}}\pm1\sigma$ (68.27\%) and
$2\sigma$ (95.45\%) limits for $N_{\text{eff}}$
obtained from the analysis of the indicated data sets.}
\begin{center}
\begin{tabular}{clccc}
\multicolumn{2}{c}{data} & $N_{\text{eff}}^{\text{gbf}}$ & $N_{\text{eff}}^{\text{mbf}}\pm1\sigma$ & $2\sigma$ \\
\hline
\multirow{3}{*}{
\begin{varwidth}{4em}
\center no\\SBL\\prior
\end{varwidth}}
	&CMB+$H_0$		&$3.84$	&$3.76^{+0.25}_{-0.23}$	&$3.29 \div 4.26$ \\
	&CMB+$H_0$+BAO		&$3.59$	&$3.71^{+0.23}_{-0.27}$	&$3.17 \div 4.18$ \\
	&CMB+$H_0$+BAO+LGC	&$3.57$	&$3.51^{+0.29}_{-0.29}$	&$3.05 \div 4.01$ \\
\hline
\multirow{4}{*}{
\begin{varwidth}{4em}
\center TH\\SBL\\prior
\end{varwidth}}
	&CMB			&$3.29$	&$3.26^{+0.21}_{-0.10}$	&$3.05 \div 3.67$ \\
	&CMB+$H_0$		&$3.23$	&$3.23^{+0.19}_{-0.12}$	&$3.05 \div 3.66$ \\
	&CMB+$H_0$+BAO		&$3.11$	&$3.23^{+0.15}_{-0.11}$	&$3.05 \div 3.55$ \\
	&CMB+$H_0$+BAO+LGC	&$3.36$	&$3.32^{+0.12}_{-0.09}$	&$3.15 \div 3.57$ \\
\hline
\multirow{4}{*}{
\begin{varwidth}{4em}
\center DW\\SBL\\prior
\end{varwidth}}
	&CMB			&$3.43$	&$3.35^{+0.16}_{-0.15}$	&$3.09 \div 3.73$ \\
	&CMB+$H_0$		&$3.19$	&$3.31^{+0.18}_{-0.13}$	&$3.08 \div 3.70$ \\
	&CMB+$H_0$+BAO		&$3.29$	&$3.30^{+0.13}_{-0.13}$	&$3.08 \div 3.60$ \\
	&CMB+$H_0$+BAO+LGC	&$3.30$	&$3.42^{+0.11}_{-0.11}$	&$3.22 \div 3.67$ \\
\hline
\end{tabular}
\end{center}
\end{table}

\begin{table*}[b]
\caption{\label{tab:meffs}
Global best-fit value $m_{s,\text{gbf}}^{\text{eff}}$,
marginal best-fit value $m_{s,\text{mbf}}^{\text{eff}}$,
$1\sigma$ (68.27\%) and $2\sigma$ (95.45\%)
intervals for $m_{s}^{\text{eff}}$
in eV
obtained from the analysis of the indicated data sets
without and
with the SBL prior
in the thermal (TH) and Dodelson-Widrow (DW) models.
We give also the corresponding quantities for $m_{s}$.}
\begin{center}
\begin{tabular}{cl|cccc|ccccc}
\multicolumn{2}{c}{data}
& $m_{s,\text{gbf}}^{\text{eff}}$ & $m_{s,\text{mbf}}^{\text{eff}}$ & $1\sigma$ & $2\sigma$ 
& $m_{s}^{\text{gbf}}$ & $m_{s}^{\text{mbf}}$ & $1\sigma$ & $2\sigma$ \\
\hline
\multirow{3}{*}{
\begin{varwidth}{4em}
\center no\\SBL\\prior
\end{varwidth}}
& CMB+$H_0$		& $0$    & $0$    & $<0.10$	   & $<0.27$          & $0$ & $0$                                           & \begin{tabular}{c}$<0.13$\\$<0.14$\end{tabular}		      & \begin{tabular}{c}$<0.38$\\$<0.44$\end{tabular}                   & \begin{tabular}{c}(TH)\\(DW)\end{tabular} \\
& CMB+$H_0$+BAO		& $0$    & $0$    & $<0.13$	   & $<0.32$          & $0$ & $0$                                           & \begin{tabular}{c}$<0.18$\\$<0.21$\end{tabular}		      & \begin{tabular}{c}$<0.51$\\$<0.65$\end{tabular}                   & \begin{tabular}{c}(TH)\\(DW)\end{tabular} \\
& CMB+$H_0$+BAO+LGC	& $0.41$ & $0.42$ & $0.28\div0.56$ & $0.15 \div 0.70$ & \begin{tabular}{c}$0.67$\\$0.79$\end{tabular} & \begin{tabular}{c}$0.62$\\$0.92$\end{tabular} & \begin{tabular}{c}$0.21\div1.14$\\$0.00\div1.11$\end{tabular} & \begin{tabular}{c}$0.00 \div 2.68$\\$0.00 \div 4.81$\end{tabular} & \begin{tabular}{c}(TH)\\(DW)\end{tabular} \\
\hline
\multirow{4}{*}{
\begin{varwidth}{4em}
\center TH\\SBL\\prior
\end{varwidth}}
& CMB			& $0.45$ & $0.42$ & $0.26 \div 0.67$ & $0.11 \div 0.89$ & $1.30$ & $1.28$ & $1.09 \div 1.36$ & $0.96 \div 1.42$ \\
& CMB+$H_0$		& $0.35$ & $0.38$ & $0.20 \div 0.61$ & $0.05 \div 0.86$ & $1.28$ & $1.28$ & $1.08 \div 1.35$ & $0.95 \div 1.40$ \\
& CMB+$H_0$+BAO		& $0.17$ & $0.37$ & $0.20 \div 0.54$ & $0.08 \div 0.75$ & $1.29$ & $1.27$ & $1.08 \div 1.35$ & $0.95 \div 1.39$ \\
& CMB+$H_0$+BAO+LGC	& $0.47$ & $0.48$ & $0.35 \div 0.60$ & $0.25 \div 0.74$ & $1.12$ & $1.27$ & $1.08 \div 1.35$ & $0.95 \div 1.40$ \\
\hline
\multirow{4}{*}{
\begin{varwidth}{4em}
\center DW\\SBL\\prior
\end{varwidth}}
& CMB			& $0.44$ & $0.36$ & $0.19 \div 0.57$ & $0.06 \div 0.83$ & $1.13$ & $1.28$ & $1.08 \div 1.35$ & $0.96 \div 1.42$ \\
& CMB+$H_0$		& $0.16$ & $0.35$ & $0.16 \div 0.53$ & $0.04 \div 0.77$ & $1.13$ & $1.28$ & $1.07 \div 1.35$ & $0.94 \div 1.39$ \\
& CMB+$H_0$+BAO		& $0.32$ & $0.28$ & $0.16 \div 0.46$ & $0.06 \div 0.64$ & $1.28$ & $1.27$ & $1.07 \div 1.34$ & $0.95 \div 1.39$ \\
& CMB+$H_0$+BAO+LGC	& $0.32$ & $0.45$ & $0.33 \div 0.58$ & $0.22 \div 0.72$ & $1.27$ & $1.28$ & $1.08 \div 1.35$ & $0.95 \div 1.40$ \\
\hline
& SBL \protect\cite{1308.5288} & & & &
& $1.27$
& $1.27$
& $1.10\div1.36$
& $0.97\div1.42$
\\
\hline
\end{tabular}
\end{center}
\end{table*}

\begin{figure*}[p]
\centering
\includegraphics*[page=1, width=\textwidth, viewport=14 14 913 913]{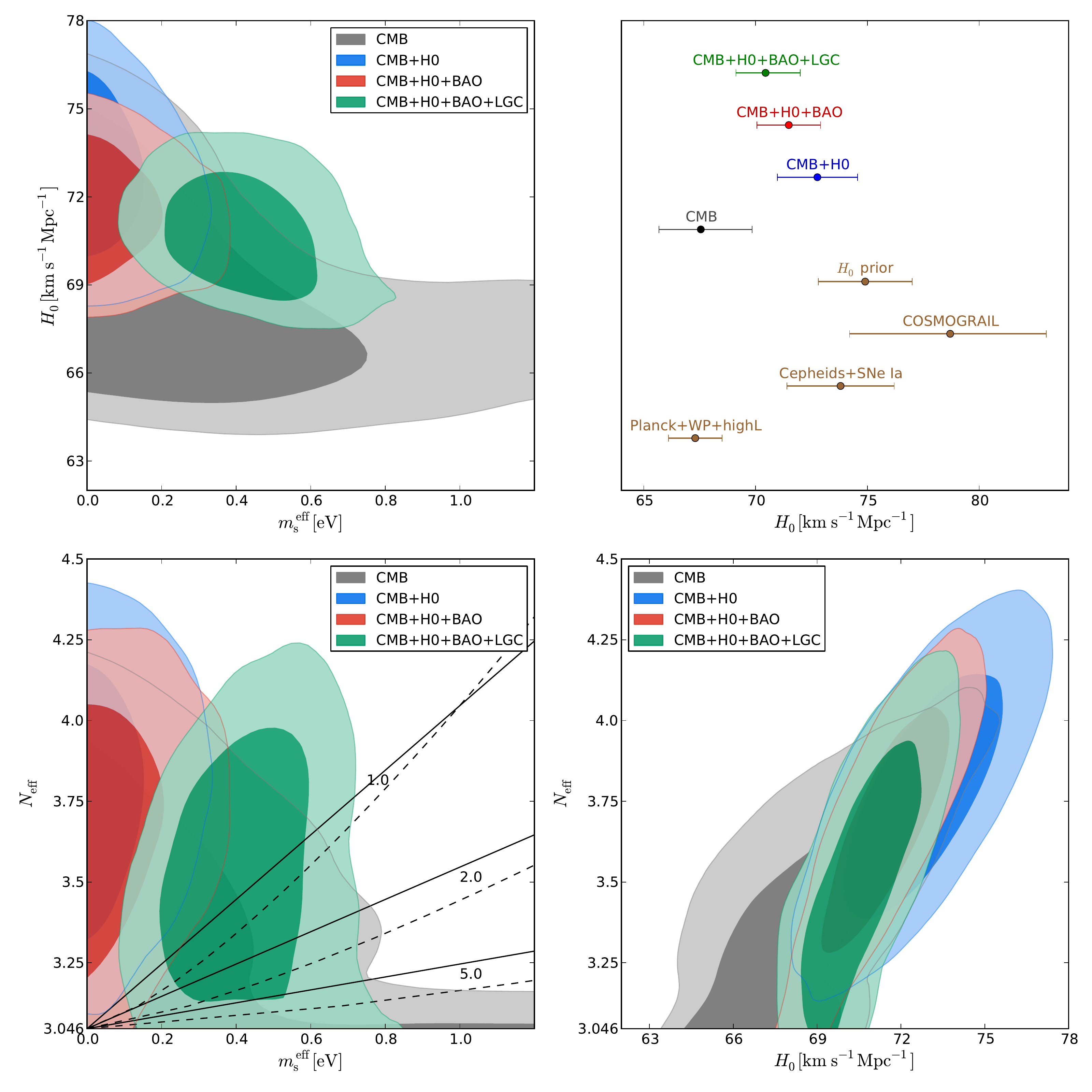}
\caption{ \label{fig:nosbl}
Results of the analysis of cosmological data alone.
The light and dark shadowed regions in the 2D plots show,
respectively, the 68\% and 95\% marginalized posterior probability regions
obtained from the analysis of the data sets indicated in the legends with corresponding color.
In the bottom-left panel $m_{s}$ is constant, with the indicated value in eV,
along the dashed lines in the thermal model
and
along the solid lines in the Dodelson-Widrow model.
The four lower intervals of $H_{0}$ in the upper-right panel correspond to:
Eq.~(\ref{H0planck}) for Planck+WP+highL \protect\cite{1303.5076},
Eq.~(\ref{H02}) for Cepheids+SNe Ia \protect\cite{1103.2976},
Eq.~(\ref{H04}) for COSMOGRAIL \protect\cite{1208.6010},
Eq.~(\ref{H0loc}) for the $H_{0}$ prior.
In all panels the labels
CMB, CMB+$H_0$, CMB+$H_0$+BAO and CMB+$H_0$+BAO+LGC
indicate the fits performed in this work.
}
\end{figure*}

\begin{figure}[t]
\centering
\includegraphics*[page=5, width=\linewidth, viewport=403 14 798 394]{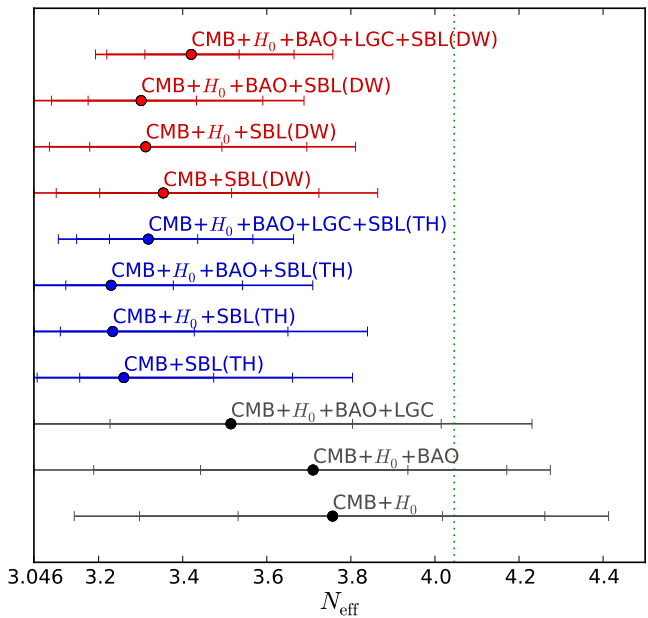}
\caption{ \label{fig:neff}
Comparison of the allowed intervals of $N_{\text{eff}}$
obtained from the fits of
CMB, CMB+$H_0$, CMB+$H_0$+BAO and CMB+$H_0$+BAO+LGC data
without (black) and with the SBL prior in the thermal (blue) and Dodelson-Widrow (red) models.
The segments in each bar correspond to 68\%, 95\% and 99\% probability.
The dotted vertical line corresponds to $\Delta N_{\text{eff}} = 1$.
}
\end{figure}

\begin{figure}[t]
\centering
\includegraphics*[page=5, width=\linewidth, viewport=2 14 397 394]{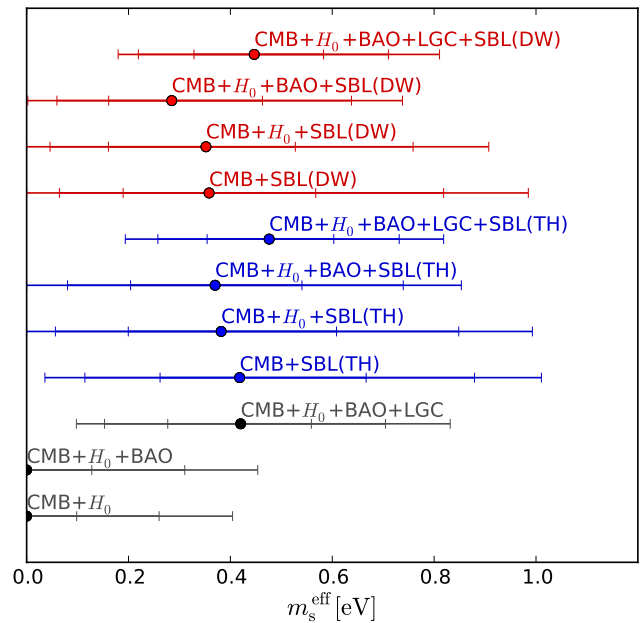}
\caption{ \label{fig:mseff}
Comparison of the allowed intervals of $m_{s}^{\text{eff}}$
obtained from the fits of
CMB, CMB+$H_0$, CMB+$H_0$+BAO and CMB+$H_0$+BAO+LGC data
without (black) and with the SBL prior in the thermal (blue) and Dodelson-Widrow (red) models.
The segments in each bar correspond to 68\%, 95\% and 99\% probability.
}
\end{figure}

\begin{figure}[t]
\centering
\includegraphics*[page=6, width=\linewidth, viewport=11 13 424 423]{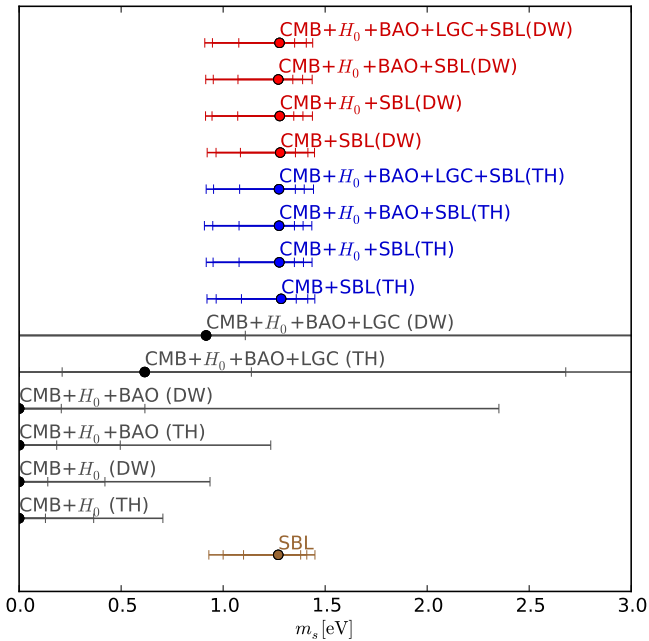}
\caption{\label{fig:ms}
Comparison of the allowed interval of $m_{s}$
obtained from the 3+1 analysis of SBL data
\cite{1308.5288}
with those obtained in the fits presented in this paper.
The segments in each bar correspond to 68\%, 95\% and 99\% probability.
The out-of-bounds upper limits obtained in the CMB+$H_0$+BAO+LGC analysis are:
$7.4 \, \text{eV}$ (99\%, TH),
$4.8 \, \text{eV}$ (95\%, DW),
$17.1 \, \text{eV}$ (99\%, DW).
}
\end{figure}

\begin{figure*}[p]
\centering
\includegraphics*[page=3, width=\textwidth, viewport=14 14 913 913]{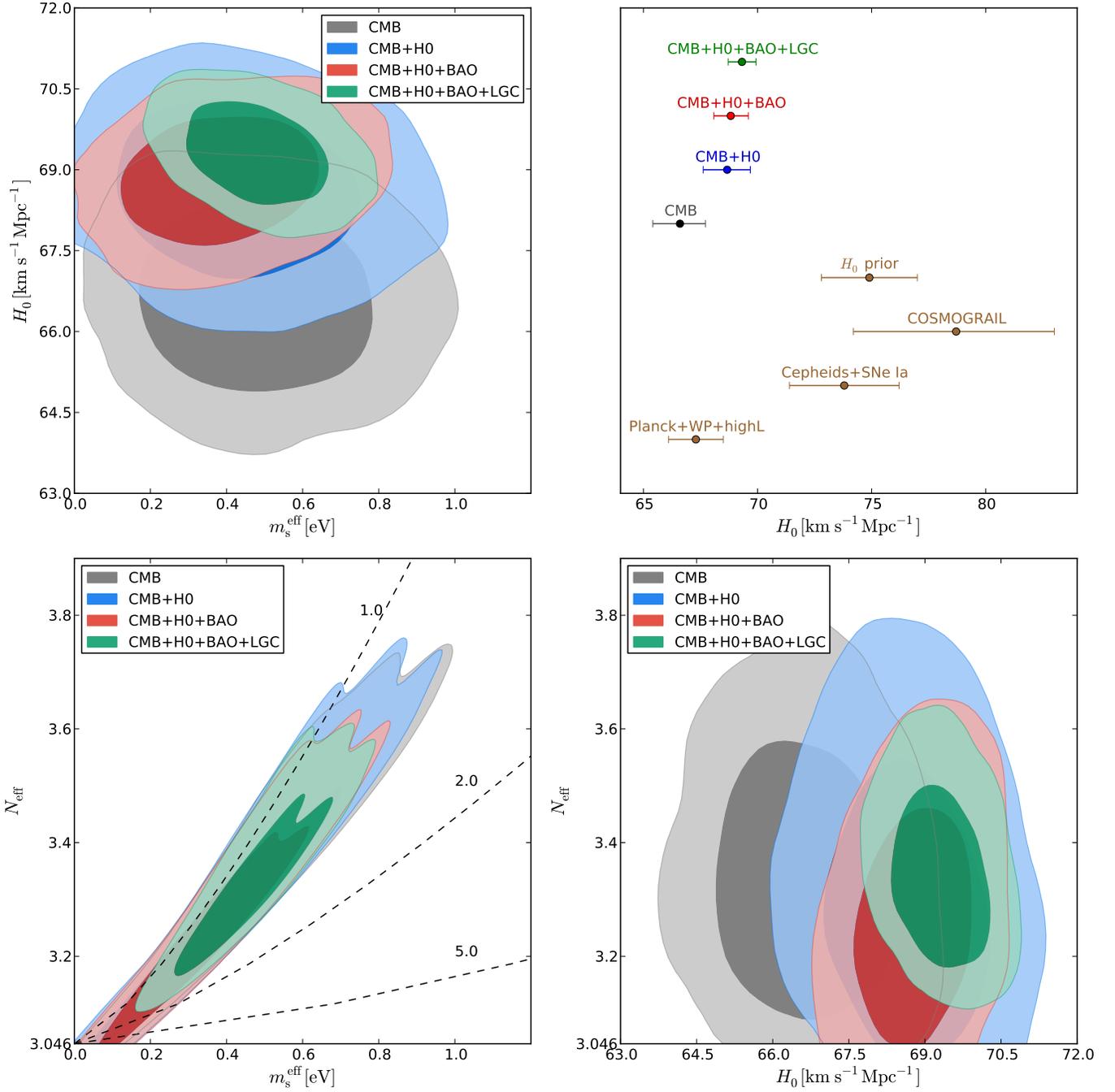}
\caption{ \label{fig:thermal}
Results of the analysis of cosmological data with the SBL prior in the thermal model.
The light and dark shadowed regions in the 2D plots show,
respectively, the 68\% and 95\% marginalized posterior probability regions
obtained from the analysis of the data sets indicated in the legends with corresponding color.
In the bottom-left panel $m_{s}$ is constant along the dashed lines,
with the indicated value in eV.
The four lower intervals of $H_{0}$ in the upper-right panel are equal to those in Fig.~\ref{fig:nosbl}.
In all panels the labels
CMB, CMB+$H_0$, CMB+$H_0$+BAO and CMB+$H_0$+BAO+LGC
indicate the fits performed in this work.
}
\end{figure*}

\begin{figure*}[p]
\centering
\includegraphics*[page=2, width=\textwidth, viewport=14 14 913 913]{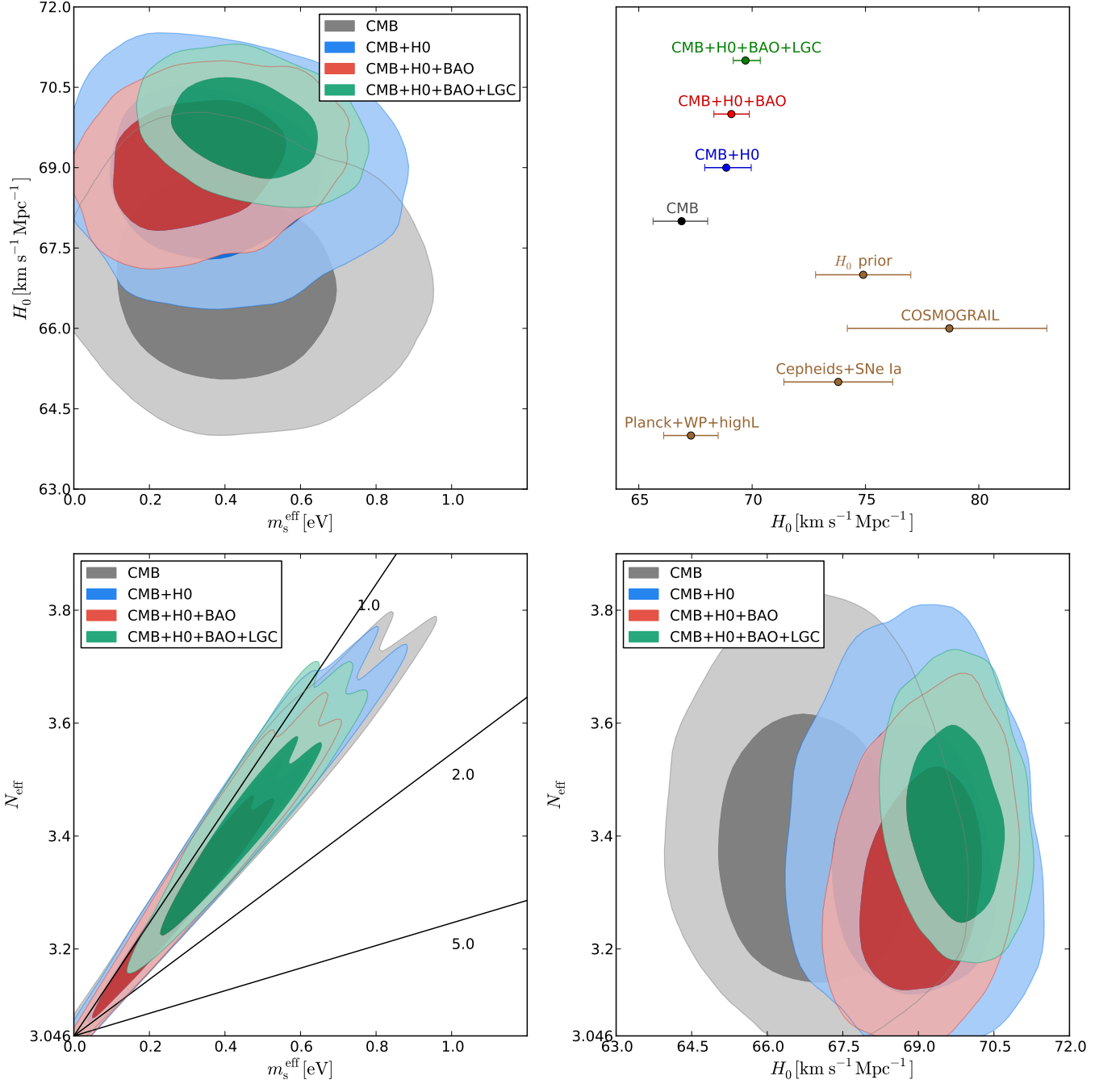}
\caption{ \label{fig:DW}
Results of the analysis of cosmological data with the SBL prior in the Dodelson-Widrow model.
The light and dark shadowed regions in the 2D plots show,
respectively, the 68\% and 95\% marginalized posterior probability regions
obtained from the analysis of the data sets indicated in the legends with corresponding color.
In the bottom-left panel $m_{s}$ is constant along the solid lines,
with the indicated value in eV.
The four lower intervals of $H_{0}$ in the upper-right panel are equal to those in Fig.~\ref{fig:nosbl}.
In all panels the labels
CMB, CMB+$H_0$, CMB+$H_0$+BAO and CMB+$H_0$+BAO+LGC
indicate the fits performed in this work.
}
\end{figure*}

\begin{figure*}[p]
\centering
\includegraphics*[page=4, width=\textwidth, viewport=14 14 913 913]{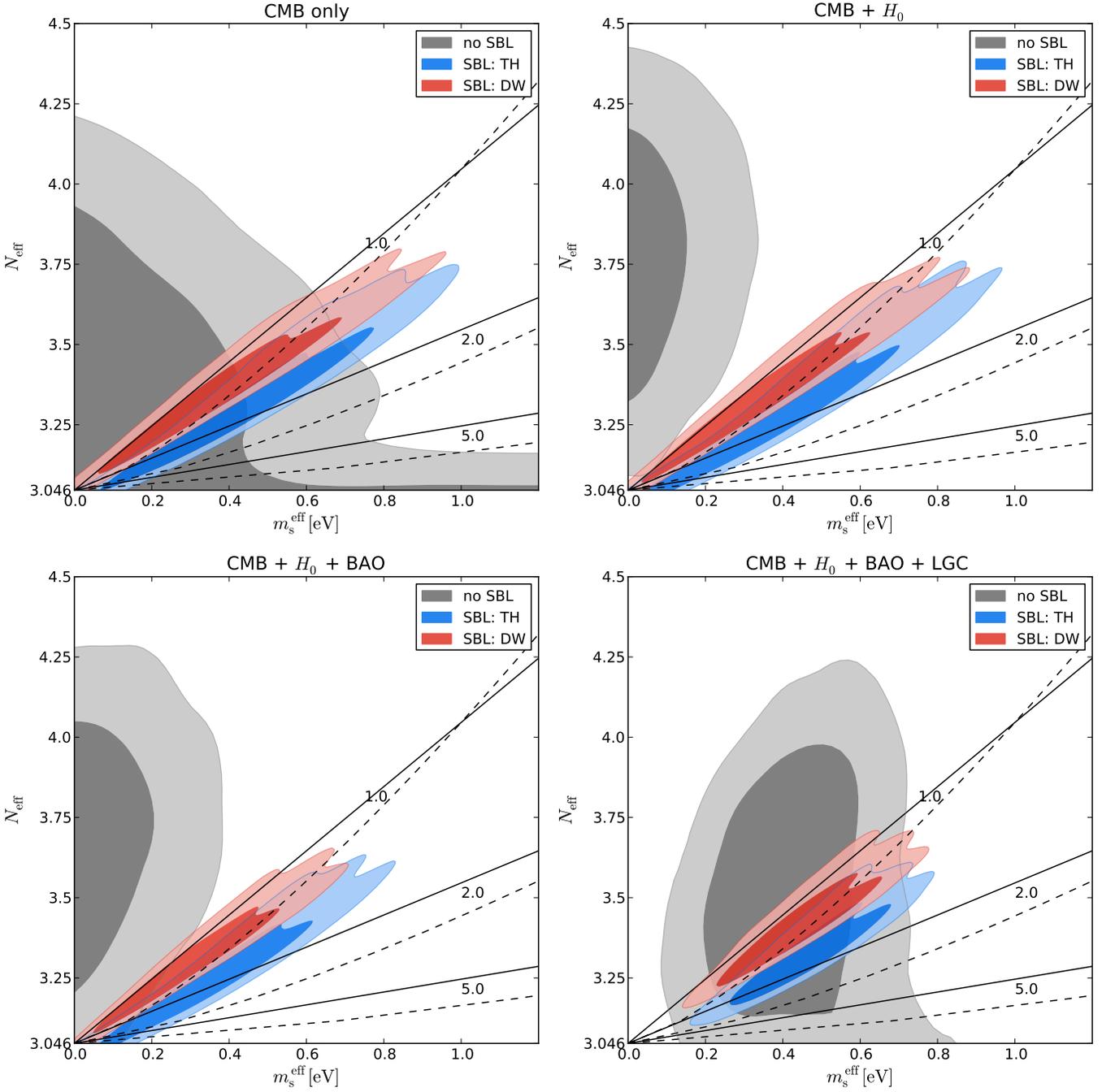}
\caption{ \label{fig:meffs}
Illustrations of the effect of the SBL prior on the results of the fits of
CMB, CMB+$H_0$, CMB+$H_0$+BAO and CMB+$H_0$+BAO+LGC data.
The value of $m_{s}$ is constant, with the indicated value in eV,
along the dashed lines in the thermal model
and
along the solid lines in the Dodelson-Widrow model.
}
\end{figure*}

\section{Cosmological Data and Local $H_0$ Measurements}
\label{Cosmological}

For our cosmological analysis we used a modified version of the publicly available software \texttt{CosmoMC}\footnote{\url{http://cosmologist.info/cosmomc/}}
\cite{astro-ph/0205436} (March 2013 version), a Monte Carlo Markov Chain (MCMC) software which computes the theoretical predictions using CAMB\footnote{\url{http://camb.info/}}
\cite{astro-ph/9911177}.
We used the following data sets and likelihood calculators:
\begin{itemize}

 \item The recent Planck data \cite{1303.5062} and likelihood codes \cite{1303.5075}
    \texttt{CamSpec}, that computes the Planck TT
    likelihood\footnote{\url{http://pla.esac.esa.int/pla/aio/planckProducts.html}}
    for the multipoles with $50\leq l\leq 2500$,
    and
    \texttt{Commander}, that computes the low-$l$ TT Planck likelihood.
 We will refer to this set as ``Planck''.
 
 \item The nine-year large-scale $E$-polarization WMAP data \cite{1212.5225}, included in the
 \texttt{CosmoMC} code through the downloadable likelihood data and
 code released by the Planck Collaboration.
 We will refer to this dataset as ``WP''.
 
 \item High-$l$ spectra from Atacama Cosmology Telescope (ACT)
 \cite{1301.1037} and 
 South Pole Telescope (SPT)
 \cite{1105.3182,1111.0932}
 and the likelihood
 code\footnote{\url{http://lambda.gsfc.nasa.gov/product/act/act_fulllikelihood_get.cfm}}
 described in \cite{1301.0776}, which is based on WMAP likelihood code \cite{1212.5226}.
 We will refer to this set as ``highL''
 and to the Planck+WP+highL dataset as ``CMB''.

 \item Baryonic Acoustic Oscillations (BAO) \cite{0910.5224} data. We used the 
 BAO measurement at  $z_{\text{eff}} = 0.2$ and $z_{\text{eff}} = 0.35$ from the Sloan Digital Sky Survey (SDSS) Data Release 7 (DR7) \cite{0812.0649} galaxy catalogue, analyzed in \cite{0907.1660} and \cite{1202.0090}, 
 BAO measurement at $z_{\text{eff}} = 0.57$ obtained from the SDSS Baryon Oscillation Spectroscopic Survey (BOSS) Data Release 9 (DR9) \cite{1207.7137} galaxy catalogue, analyzed in \cite{1203.6594}, and 
 the measurement at $z_{\text{eff}}=0.1$ obtained in \cite{1106.3366} using data from the 6dF Galaxy Survey (6dFGS) \cite{0903.5451}.
 We will refer to this set as ``BAO''.

 \item Local Galaxy Cluster data from the Chandra Cluster Cosmology Project \cite{0812.2720} 
 observations, from which the cluster mass distribution at low and high redshift is calculated.
 The likelihood has been presented in \cite{1202.2889} 
 and a \texttt{CosmoMC} module is publicly available\footnote{ \url{http://hea.iki.rssi.ru/400d/cosm/} }.
 We will refer to this set as ``LGC''.

\end{itemize}

The analysis of Planck+WP+highL data
performed by the Planck collaboration
in the framework of the standard $\Lambda$CDM cosmological model
gave for the Hubble constant the value
(see Eq.~(51) of Ref.~\cite{1303.5076})
\begin{equation}
H_{0} = 67.3 \pm 1.2 \, \text{km} \, \text{s}^{-1} \, \text{Mpc}^{-1}
\,.
\label{H0planck}
\end{equation}
This value has a remarkable tension with the results of
recent direct local astrophysical measurements of $H_{0}$,
which found significantly higher values.
Here we consider the following two compatible measurements:

\begin{description}

\item[Cepheids+SNe Ia]
Hubble Space Telescope (HST)
observations of Cepheid variables in the host galaxies of eight SNe Ia
have been used to calibrate the supernova magnitude-redshift relation,
leading to
\cite{1103.2976}
\begin{equation}
H_0 = 73.8 \pm 2.4 \, \text{km} \, \text{s}^{-1} \, \text{Mpc}^{-1}
\,.
\label{H02}
\end{equation}
This value is in agreement with the result
$H_0 = 74.3 \pm 2.6 \, \text{km} \, \text{s}^{-1} \, \text{Mpc}^{-1}$
obtained in the Carnegie Hubble Program (Carnegie HP)
\cite{1208.3281}
through a
recalibration of the secondary distance methods used in the HST Key Project.
We do not use the Carnegie Hubble Program
value of $H_0$,
because there is an overlap between the HST and Carnegie HP
sets of SNe Ia data
that induces a correlation in the statistical part of the uncertainty
which is unknown to us.

\item[COSMOGRAIL]
Strong gravitational lensing time delay measurements of
the system RXJ1131-1231,
observed as part of the
COSmological MOnitoring of GRAvitational Lenses (COSMOGRAIL) project,
led to
\cite{1208.6010}:
\begin{equation}
H_0 = 78.7 \pm 4.5 \, \text{km} \, \text{s}^{-1} \, \text{Mpc}^{-1}
\,.
\label{H04}
\end{equation}

\end{description}

Combining these local astrophysical measurements of $H_0$,
we obtained the local weighted average
\begin{equation}
H_0 = 74.9 \pm 2.1 \, \text{km} \, \text{s}^{-1} \, \text{Mpc}^{-1}
\,.
\label{H0loc}
\end{equation}
Since this value differs from
that in Eq.~(\ref{H0planck}) by about $3.1\sigma$,
there is a tension between the CMB and local determinations of $H_0$.
One can see this tension also from the graphical representation of
Eqs.~(\ref{H0planck})--(\ref{H0loc})
in the upper-right panel of Fig.~\ref{fig:nosbl}.

Let us emphasize however
that the Planck value of $H_0$ in Eq.~(\ref{H0planck})
has been obtained assuming the standard $\Lambda$CDM cosmological model
in which $N_{\text{eff}}$ is assumed to have the SM value $N_{\text{eff}}^{\text{SM}}$.
Since $H_0$ and $N_{\text{eff}}$ are positively correlated
(see Refs.~\cite{1104.2333,1307.0637}),
one can expect that leaving $N_{\text{eff}}$ free the tension is reduced.
Indeed, from the analysis of CMB data
without constraints on $N_{\text{eff}}$ the Planck collaboration obtained\footnote{
See page 148
of the tables with 68\% limits available at
\url{http://www.sciops.esa.int/wikiSI/planckpla/index.php?title=Cosmological_Parameters&instance=Planck_Public_PLA}.
}
\begin{equation}
H_0 = 69.7 \pm 2.8 \, \text{km} \, \text{s}^{-1} \, \text{Mpc}^{-1}
\,.
\label{H0CMB}
\end{equation}
Hence,
the tension of CMB Planck data with the local weighted average value of $H_0$ in Eq.~(\ref{H0loc})
almost disappears when $N_{\text{eff}}$ is not constrained to its SM value.
Hence,
we performed a fit of cosmological data
with a Gaussian prior for $H_0$
having mean and standard deviation given by the local average in Eq.~(\ref{H0loc}).
In the following we refer to this prior as ``$H_0$''.

Since we are interested in studying the effect on the analysis of cosmological data
of a sterile neutrino mass motivated by SBL oscillation anomalies,
we consider an extension of the standard cosmological model
in which both $N_{\text{eff}}$ and $m_{s}^{\text{eff}}$
are free parameters to be determined by the fit of the data.

Figure~\ref{fig:nosbl}
and
the first parts of
Tabs.~\ref{tab:H0}, \ref{tab:neff} and \ref{tab:meffs}
shows the results for
$H_0$,
$N_{\text{eff}}$ and
$m_{s}^{\text{eff}}$
obtained from the fits of
CMB, CMB+$H_0$, CMB+$H_0$+BAO and CMB+$H_0$+BAO+LGC data.
In Tab.~\ref{tab:meffs}
we give also the corresponding results for
$m_{s} \simeq m_{4}$,
which depends on the statistical distribution of sterile neutrinos.
Therefore,
we distinguish the results for
$m_{s}$ obtained in the
thermal (TH) and Dodelson-Widrow (DW) models
using, respectively,
Eqs.~(\ref{THmeff}) and (\ref{DWmeff}).
In Figs.~\ref{fig:neff}, \ref{fig:mseff} and \ref{fig:ms}
we compare graphically the allowed ranges of $N_{\text{eff}}$, $m_{s}^{\text{eff}}$ and $m_{s}$
obtained in the different fits.

From the bottom-left panel in Fig.~\ref{fig:nosbl},
one can see that the fit of CMB data alone restricts
$m_{s}^{\text{eff}}$ to small values only for
$N_{\text{eff}}\gtrsim3.2$,
whereas there is a tail of allowed large values of
$m_{s}^{\text{eff}}$
for smaller $N_{\text{eff}}$.
This is in agreement with Fig.~28-right of Ref.~\cite{1303.5076},
where it has been explained as corresponding to the case in which the sterile neutrino
behaves as warm dark matter,
because its mass is large and it becomes non-relativistic well before recombination.
This happens in both the thermal and Dodelson-Widrow models,
as one can infer from Eqs.~(\ref{THmeff}) and (\ref{DWmeff}).
The presence of this tail of the posterior distribution of
$m_{s}^{\text{eff}}$
implies that the posterior distributions of the fitted parameters
depend on the arbitrary upper value chosen for $m_{s}^{\text{eff}}$
in the \texttt{CosmoMC} runs
(we chose $m_{s}^{\text{eff}}<5\,\text{eV}$,
whereas the Planck Collaboration chose
$m_{s}^{\text{eff}}<3\,\text{eV}$).
Hence, we do not present in the tables the numerical results of the fit of CMB data alone,
which suffer from this arbitrariness.

The addition of the local $H_0$ prior leads to an increase of
$N_{\text{eff}}$
which evicts the
large-$m_{s}^{\text{eff}}$ and small-$N_{\text{eff}}$
region in which the sterile neutrino
behaves as cold dark matter.
This can be seen from the
CMB+$H_0$
allowed regions in Fig.~\ref{fig:nosbl}
and the corresponding upper limits for
$m_{s}^{\text{eff}}$
and
$m_{s}$
in Figs.~\ref{fig:mseff} and \ref{fig:ms} and in Tab.~\ref{tab:meffs}.
The further addition of BAO data slightly lowers the
best-fit values and allowed ranges of
$H_0$
and
$N_{\text{eff}}$
(see Figs.~\ref{fig:nosbl} and \ref{fig:neff}
and
Tabs.~\ref{tab:H0} and \ref{tab:neff}).
Hence,
the upper limits for
$m_{s}^{\text{eff}}$
and
$m_{s}$
in Figs.~\ref{fig:mseff} and \ref{fig:ms} and in Tab.~\ref{tab:meffs}
are slightly larger,
but still rather stringent,
of the order of
$m_{s}^{\text{eff}} \lesssim 0.3 \, \text{eV}$
and
$m_{s} \lesssim 0.6 \, \text{eV}$
at $2\sigma$.

Comparing the
CMB+$H_0$
and
CMB+$H_0$+BAO
allowed intervals of
$m_{s}$
in Tab.~\ref{tab:meffs} and Fig.~\ref{fig:ms}
with that obtained from the analysis of SBL data
in the framework of 3+1 mixing \cite{1308.5288},
it is clear that there is a tension\footnote{
Possible ways of solving this tension have been discussed recently,
but before the Planck data release, in
Refs.~\cite{1112.4661,1203.6828,1212.1689}.
}:
about
$5.0\sigma$,
$4.6\sigma$,
$4.1\sigma$,
$3.5\sigma$,
respectively,
in the
CMB+$H_0$(TH),
CMB+$H_0$(DW),
CMB+$H_0$+BAO(TH)
CMB+$H_0$+BAO(DW)
fits.
The tensions are smaller in the Dodelson-Widrow model
and this could be an indication in favor of this case if
SBL oscillations will be confirmed by future experiments
(see Refs.~\cite{1107.2335,1204.5379,1205.4419,1304.2047,1304.7721,1307.7097,1308.5700,1308.6822}).

Let us now consider the inclusion of the LGC data set
in the cosmological fit.
As discussed in Ref.~\cite{1307.7715},
the measured amount of clustering of galaxies
\cite{1202.2889,0812.2720}
is smaller than that
obtained by evolving the primordial density fluctuations
with the relatively large matter density at recombination
measured precisely by Planck
\cite{1303.5076}.
The correlation of a relatively large
matter density and clustering of galaxies
can be quantified through the approximate relation
$
\sigma_{8}
\propto
\Omega_{m}^{0.563}
$
\cite{astro-ph/0312395,astro-ph/0407158}
which relates the rms amplitude
$\sigma_{8}$
of linear fluctuations today at a scale of $8 h^{-1} \, \text{Mpc}$
(where $h = H_{0} / 100 \, \text{km} \, \text{s}^{-1} \, \text{Mpc}^{-1}$)
with the present matter density
$\Omega_{m}$.
As discussed in Ref.~\cite{1307.7715},
the value of
$\sigma_{8}$ and the amount of clustering of galaxies
can be lowered by adding to the $\Lambda$CDM cosmological model
hot dark matter in the form of sterile neutrinos with eV-scale masses\footnote{
Let us note that there was already a tension between
LGC data and pre-Planck CMB data
and the sterile neutrino solution
was proposed in Refs.~\cite{1202.2889,1301.4791}}.
The free-streaming of these sterile neutrinos suppresses the growth of structures
which are smaller than the free-streaming length,
leading to a suppression of $\sigma_{8}$
with respect to the $\Lambda$CDM approximate relation
$
\sigma_{8}
\propto
\Omega_{m}^{0.563}
$.
In this way,
the relatively large Planck value of $\Omega_{m}$
can be reconciled
with the relatively small amount of local galaxy clustering in the LGC data set
and the corresponding relatively small value of $\sigma_{8}$.

Hence,
the inclusion of LGC data in the cosmological fits
favors the existence of a sterile neutrino with a mass of the order of that
required by SBL data,
which is at least partially thermalized in the early Universe
\cite{1307.7715}.
The results of our
CMB+$H_0$+BAO+LGC
fit given in Figs.~\ref{fig:nosbl}, \ref{fig:neff}, \ref{fig:mseff}, \ref{fig:ms}
and
Tabs.~\ref{tab:H0}, \ref{tab:neff}, \ref{tab:meffs}
confirm this expectation.
In particular,
from the allowed intervals of
$m_{s}$
in Tab.~\ref{tab:meffs} and Fig.~\ref{fig:ms}
one can see that the tension between cosmological data and SBL 3+1 oscillations
disappears with the inclusion of LGC data.

In the following section we analyze the cosmological data
using as prior distribution for $m_s$
the distribution obtained from the analysis of SBL data.
This is perfectly consistent
in the case of
CMB+$H_0$+BAO+LGC
cosmological data.
However,
we present also the results obtained with the
CMB,
CMB+$H_0$ and
CMB+$H_0$+BAO
cosmological data,
in spite of the tension with SBL data discussed above,
because
we think that one cannot dismiss the results of laboratory experiments
on the basis of cosmological observations,
whose interpretation has larger uncertainties.

\section{SBL Prior}
\label{SBL}

The existence of light sterile neutrinos has been considered in recent years as a plausible possibility
motivated by the measurements of anomalies which can be explained by short-baseline (SBL) neutrino oscillations
generated by a squared-mass difference of the order of $1 \, \text{eV}^2$:
the reactor anomaly
\cite{1101.2663,1101.2755,1106.0687},
the Gallium anomaly
\cite{1006.3244,1210.5715}
and
the LSND anomaly
\cite{hep-ex/0104049}.
Here we consider the results of the analysis of SBL data
in the framework of 3+1 mixing presented in Ref.~\cite{1308.5288}.
Following Refs.~\cite{1207.6515,1302.6720,1303.4654},
we use the posterior distribution of
$m_s \simeq m_4 \simeq \sqrt{\Delta{m}^2_{41}}$
obtained from the analysis of SBL data
as a prior in the \texttt{CosmoMC} analysis of cosmological data.
The range of $m_{s}$ allowed by the analysis of SBL data \cite{1308.5288}
is shown in Fig.~\ref{fig:ms} and Tab.~\ref{tab:meffs}.
Note that the SBL prior on $m_s$ has different cosmological implications
in the thermal and Dodelson-Widrow models,
because the
$\Delta N_{\text{eff}}$ dependence of the effective mass
$m_{s}^{\text{eff}}$
is different (see Eqs.~(\ref{THmeff}) and (\ref{DWmeff})).

Figure~\ref{fig:thermal} shows the results of the analysis of
CMB, CMB+$H_0$, CMB+$H_0$+BAO and CMB+$H_0$+BAO+LGC data
with the SBL prior in the thermal model.
For convenience,
the effect of the SBL prior
on the allowed regions in the
$m_{s}^{\text{eff}}$--$N_{\text{eff}}$ plane
is illustrated clearly in Fig.~\ref{fig:meffs},
where each panel shows the change of the allowed regions due to the inclusion of the SBL prior
corresponding to the analysis of the indicated data set.
One can see that in all four analyses the SBL prior forces the
allowed region in an area near the dashed line which corresponds to
$m_{s} = 1 \, \text{eV}$.
In order to keep $m_{s}$ at the eV scale without increasing too much $m_{s}^{\text{eff}}$,
which is forbidden by the cosmological data,
$N_{\text{eff}}$
is forced towards low values.

In the case of the CMB+$H_0$+BAO+LGC cosmological data set
the addition of the SBL prior approximately confirms the allowed range of
$m_{s}^{\text{eff}}$
(see Fig.~\ref{fig:mseff} and Tab.~\ref{tab:meffs}),
but requires a lower
$N_{\text{eff}}$
(see Fig.~\ref{fig:neff} and Tab.~\ref{tab:neff}),
which must be smaller than about 3.7 with 99\% probability.
As discussed in Ref.~\cite{1303.5368},
in the standard cosmological scenario
active-sterile neutrino oscillations
generated by values of the mixing parameters allowed by the fit of SBL data
imply $\Delta N_{\text{eff}} = 1$.
Therefore,
it is likely that the compatibility of
the neutrino oscillation explanation of the SBL anomalies
with cosmological data
requires that active-sterile neutrino oscillations in the early Universe
are somewhat suppressed by a non-standard mechanism,
as, for example,
a large lepton asymmetry
\cite{1204.5861,1206.1046,1302.1200,1302.7279}.

As one can see from
Figs.~\ref{fig:neff}, \ref{fig:mseff}, \ref{fig:DW} and \ref{fig:meffs}
and from
Tabs.~\ref{tab:neff} and \ref{tab:meffs},
similar conclusions are reached in the Dodelson-Widrow model.
One can note, however, that in this case
slightly larger values of $N_{\text{eff}}$
are allowed with respect to the thermal case,
and there is a slightly better compatibility
of cosmological and SBL data.
This happens because for a given value of $m_{s}$ given mainly by SBL data
and an upper bound on
$m_{s}^{\text{eff}}$ given by cosmological data
slightly larger values of
$\Delta N_{\text{eff}} \leq 1$ are allowed by Eq.~(\ref{DWmeff}) in the Dodelson-Widrow model
than by
Eq.~(\ref{THmeff}) in the thermal model.

\section{Conclusions}
\label{Conclusions}

In this paper we have analyzed the most recent cosmological data,
including those of the Planck experiment \cite{1303.5062,1303.5076},
taking into account the possible existence of a sterile neutrino with a mass $m_s$ in the eV range,
which could have the effect of dark radiation in the early Universe.
We investigated three effects:
1) the contribution of local measurements of the Hubble constant $H_0$;
2) the effect of the measurements of the mass distribution of local galaxy clusters \cite{1307.7715};
3) the assumption of a prior distribution for $m_s$
obtained from the analysis of short-baseline oscillation data
in the framework of 3+1 mixing, which requires a sterile neutrino mass
between about 0.9 and 1.5 eV \cite{1308.5288}.
For the statistical distribution of the sterile neutrinos
we considered the two most studied cases:
the thermal model and the Dodelson-Widrow model \cite{hep-ph/9303287}.

We have shown that the local measurements of the Hubble constant $H_0$
induce an increase of the value of the effective number of relativistic degrees of freedom
$N_{\text{eff}}$
above the Standard Model value.
This is an indication in favor of the existence of sterile neutrinos
and their contribution to dark radiation.
However,
we obtained that the sterile neutrino mass has a $2\sigma$ upper bound of
about 0.5 eV in the thermal model
and
about 0.6 eV in the Dodelson-Widrow model.
Hence, there is a tension
between cosmological and SBL data.
The Dodelson-Widrow model is slightly more compatible with SBL data
and it may turn out that it is favorite if
SBL oscillations will be confirmed by future experiments\footnote{
The existence of sterile neutrinos with eV-scale masses
can be tested also in
$\beta$-decay
\cite{Riis:2010zm,SejersenRiis:2011sj,Formaggio:2011jg,1203.2632,1210.5715,1212.3805}
and
neutrinoless double-$\beta$ decay experiments
\cite{Barry:2011wb,1110.5795,1111.1069,1206.2560,1210.5715,1212.3805,1308.5802}.
}
(see Refs.~\cite{1107.2335,1204.5379,1205.4419,1304.2047,1304.7721,1307.7097,1308.5700,1308.6822}).

The tension between cosmological and SBL data
disappears if we consider also the measurements of the local galaxy cluster mass distribution,
which favor the existence of sterile neutrinos with eV-scale masses
which can suppress the small-scale clustering of galaxies
through free-streaming \cite{1307.7715}.
In this case we obtained a cosmologically allowed range for the sterile neutrino mass
which at $2\sigma$ can be as large as about
2.7 eV in the thermal model
and
4.8 eV in the Dodelson-Widrow model.

In the combined fit of
cosmological and SBL data
the sterile neutrino mass is restricted around 1 eV by the SBL prior
and the cosmological limits on the sterile neutrino mass
$m_{s}^{\text{eff}}$
imply that the contribution of the sterile neutrino
to the effective number of relativistic degrees of freedom
$N_{\text{eff}}$
is likely to be smaller than one.
In this case,
the production of sterile neutrinos in the early Universe
must be somewhat suppressed by a non-standard mechanism,
as, for example,
a large lepton asymmetry
\cite{1204.5861,1206.1046,1302.1200,1302.7279}.
The slightly smaller suppression required by the
Dodelson-Widrow model and the slightly better compatibility
of cosmological and SBL data in this model
may be indications in its favor,
with respect to the thermal model.


\end{document}